\documentclass[12pt]{article}
\usepackage{fullpage,citesort,
epsfig,graphics,amsbsy,amssymb,amsmath
}
\usepackage{caption}
\newcommand{\beq}{\begin{equation}}
\newcommand{\eeq}{\end{equation}}
\newcommand{\bea}{\begin{eqnarray}}
\newcommand{\eea}{\end{eqnarray}}

\newcommand{\Tr}{{\rm Tr}}

\newcommand{\ket}[1]{|#1\rangle}

\def\math{\mathsurround=0pt }
\def\leftrightarrowfill{$\math \mathord\leftarrow \mkern-6mu
 \cleaders\hbox{$\mkern-2mu \mathord- \mkern-2mu$}\hfill
 \mkern-6mu \mathord\rightarrow$}
\def\overleftrightarrow#1{\vbox{\ialign{##\crcr
     \leftrightarrowfill\crcr\noalign{\kern-1pt\nointerlineskip}
     $\hfil\displaystyle{#1}\hfil$\crcr}}}

\let\l=\lambda

 \def\bd{\begin{document}} \def\ed{\end{document}}
\def\ds{\documentstyle} \let\fr=\frac \let\bl=\bigl \let\br=\bigr
\let\Br=\Bigr \let\Bl=\Bigl
\let\bm=\bibitem
\let\na=\nabla
\let\pa=\partial \let\ov=\overline
\def\ft#1#2{{\textstyle{{\scriptstyle #1}\over {\scriptstyle #2}}}}
\def\fft#1#2{{#1 \over #2}}
\def\vp{\varphi}
\def\sst#1{{\scriptscriptstyle #1}}
\def\oneone{\rlap 1\mkern4mu{\rm l}}
\def\td{\tilde}
\def\wtd{\widetilde}
\def\dalemb#1#2{{\vbox{\hrule height .#2pt
        \hbox{\vrule width.#2pt height#1pt \kern#1pt
                \vrule width.#2pt}
        \hrule height.#2pt}}}
\def\square{\mathord{\dalemb{6.8}{7}\hbox{\hskip1pt}}}
\def\wtd{\widetilde}
\def\R{\rlap{\rm I}\mkern3mu{\rm R}}
\def\im{{\rm i}}
\def\tilg{\tilde{g}}
\def\tilF{\tilde{F}}
\def\tilA{\tilde{A}}
\def\varf{\varphi}
\def\tilf{\tilde{\phi}}
\def\tilh{\tilde{h}}
\def\rme{{\rm e}}
\def\ep{\epsilon}
\def\0{{(0)}}
\def\9{{(9)}}
\def\8{{(8)}}
\def\7{{(7)}}
\def\6{{(6)}}
\def\5{{(5)}}
\def\4{{(4)}}
\def\3{{(3)}}
\def\2{{(2)}}
\def\1{{(1)}}
\newcommand{\trace}{{\rm Tr}}
\newcommand{\ub}{\overline{U}}
\newcommand{\vb}{\overline{V}}
\newcommand{\uh}{\widehat{U}}
\newcommand{\vh}{\widehat{V}}
\newcommand{\ubh}{\overline{\widehat{U}}}
\newcommand{\vbh}{\overline{\widehat{V}}}
\newcommand{\lb}{\bar{\l}}
\newcommand{\Fb}{\overline{F}}
\newcommand{\Fh}{\widehat{F}}
\newcommand{\Fbh}{\overline{\widehat{F}}}
\newcommand{\Ab}{\overline{A}}
\newcommand{\Ah}{\widehat{A}}
\newcommand{\Abh}{\overline{\widehat{A}}}
\newcommand{\Gb}{\overline{G}}
\newcommand{\Gh}{\widehat{G}}
\newcommand{\Gbh}{\overline{\widehat{G}}}
\newcommand{\Pb}{\overline{P}}
\newcommand{\Ph}{\widehat{P}}
\newcommand{\Pbh}{\overline{\widehat{P}}}
\newcommand{\Qb}{\overline{Q}}
\newcommand{\Qh}{\widehat{Q}}
\newcommand{\Qbh}{\overline{\widehat{Q}}}
\newcommand{\Bb}{\overline{B}}
\newcommand{\Bh}{\widehat{B}}
\newcommand{\Bbh}{\overline{\widehat{B}}}
\newcommand{\fhns}{\hat{F}^{\rm (NS)}}
\newcommand{\fhrr}{\hat{F}^{\rm (RR)}}
\newcommand{\ahns}{\hat{A}^{\rm (NS)}}
\newcommand{\ahrr}{\hat{A}^{\rm (RR)}}
\newcommand{\hhrr}{\hat{H}^{\rm (RR)}}
\newcommand{\hchi}{\hat{\chi}}
\newcommand{\hphi}{\hat{\phi}}
\newcommand{\htau}{\hat{\tau}}
\newcommand{\cG}{{\cal G}}
\newcommand{\cGb}{\overline{{\cal G}}}
\newcommand{\cH}{{\cal H}}
\newcommand{\cP}{{\cal P}}
\newcommand{\cPb}{\overline{{\cal P}}}
\newcommand{\cQ}{{\cal Q}}
\newcommand{\cQb}{\overline{{\cal Q}}}
\newcommand{\cM}{{\cal M}}
\newcommand{\cN}{{\cal N}}
\newcommand{\cO}{{\cal O}}
\newcommand{\cD}{{\cal D}}
\newcommand{\cL}{{\cal L}}
\newcommand{\vpp}{\mbox{$\langle{\scriptstyle++}\rangle$}}
\newcommand{\vmp}{\mbox{$\langle{\scriptstyle-+}\rangle$}}
\newcommand{\vppp}{\mbox{$\langle{\scriptstyle+++}\rangle$}}
\newcommand{\vmpp}{\mbox{$\langle{\scriptstyle-++}\rangle$}}
\newcommand{\vpmp}{\mbox{$\langle{\scriptstyle+-+}\rangle$}}
\begin{document}
\setlength{\captionmargin}{36pt}
\setlength{\captionmargin}{36pt}
\begin{titlepage}
\begin{flushright}
\phantom{UFIFT-HEP}
\end{flushright}
\vskip 3cm
\begin{center}
\begin{large}
  {\bf Internal and Super Symmetry in String Bit Models}
\end{large}

\vskip 2cm
{\large
 Charles B. Thorn\footnote{E-mail  address: {\tt thorn@phys.ufl.edu}}
}
\vskip0.20cm
{\it Institute for Fundamental Theory,\\
Department of Physics, University of Florida,
Gainesville FL 32611}
\item

\vskip 1.0cm
\end{center}

\begin{abstract}
\noindent  We study in a general way the construction of string bit Hamiltonians which
are supersymmetric, We construct several  quadratic and quartic polynomials
in string bit creation and annihilation operators
${\bar\phi}^A_{a_1\cdots a_n}$,  ${\phi}^A_{a_1\cdots a_n}$,
  which commute with the supersymmetry generators $Q^a$. Among these operators
  are ones with the spinor tensor structure required
  to provide   the lightcone worldsheet vertex insertion
  factors needed to give the
  correct interactions for the IIB superstring, whenever a closed string
  separates into two closed strings or two closed strings join into one.

\end{abstract}
\vfill
\end{titlepage}
\section{Introduction}
The aim of string bit models \cite{thornsakh} is to provide a {\it bona fide}
quantum mechanical
system whose dynamics leads, in a certain limit, to string theory.
Since most consistent versions of string theory in a Poincar\'e 
invariant background
contain a massless spin 2 particle, a successful string bit model could
also provide a {\it bona fide} quantum theory of gravity.

The central premise underlying string bit models is that spacetime is,
at least in part, an emergent concept. This idea has close parallels to
't Hooft's proposal that the world is a hologram \cite{thoofthologram}. 
String bit models developed from an
early proposal to define the worldsheet system of string theory 
on a lightcone lattice:
Choose lightcone parameterization \cite{goddardgrt,mandelstamlc}, $\tau=x^+$,
${\cal P}^+=1$, where for a Lorentz vector $V^\mu$, 
$V^\pm=(V^0\pm V^1)/\sqrt{2}$
and ${\cal P}^\mu d\sigma$ is the Lorentz momentum contained in the element
$d\sigma$ of string, and put
the resultant system on a rectangular $\sigma,\tau$ lattice \cite{gilest}.
Because of the way $\sigma$ has been fixed, each lattice site is 
assigned exactly one unit $m$ of $P^+$.
Then a little reflection shows that the lattice system, after taking
the continuum limit of the time parameter,
defines a quantum system of $M=P^+/m$ particles (string bits), ordered 
on a line,  moving in the transverse space only. Their dynamics is
Galilei invariant with  $m$ playing the
role of Newtonian mass. These simple observations already show that
$P^+$ is emergent: it is simply the total Newtonian mass of a chain
of a very large number of string bits. Its conjugate $x^-$ ``emerges''
from the string bit dynamics.

However, there is an aspect of the story just told that is artificial. 
The ordering of the string bits on
a chain is by {\it fiat}. If the string bits are indeed independent
fundamental entities, there should be interactions between all pairs
of bits, not just the nearest neighbors of bits ordered on a chain.
In other words, we should be able
to describe the presence of a string bit in the
language
of second quantization by the action of a creation
operator $\phi^\dagger$ on the empty state $\ket{0}$, with the
dynamics given by the Hamiltonian
as a function of creation and annihilation operators. 't Hooft has
identified a way to enhance chain ordering by taking a large $N$ limit
\cite{thooftlargen}.
Promote the bit annihilation operator to an $N\times N$ matrix operator
$\phi_\alpha^\beta$. Then a Hamiltonian with the structure $N^{-1}\Tr 
\phi^\dagger \phi^\dagger\phi\phi$
describes chains of bits with nearest neighbor interactions in the
limit $N\to\infty$ \cite{thornfock}. Taking $N$ large rigs the dynamics
in favor of chain formation. But if the nearest neighbor interactions
are not sufficient to make a long chain of bits stable against decay into 
smaller chains, continuous strings would not form. Indeed, the
tachyonic instability of the bosonic string can be traced to such
a failure.

As shown by the stability of superstring \cite{rns,gso,greenschwarz}, one cure for such an instability
is to include a mix of bosonic and fermionic string bits \cite{bergmantsubit}.
In recent work on these models 
\cite{sunthorn,thornspace,thornprotobits,chens,chen,songge} 
and in this article,
we adorn the string bit annihilation operator with three types of indices
\bea
(\phi_{[a_1\cdots a_n]}^A)_\alpha^{\ \beta}, 
\label{bitann}
\eea
The Greek letters $\alpha,\beta=1,\ldots,N$ are color indices labeling
the matrix elements of the matrix operators, whose function is to provide
a way to favor nearest neighbor chain interaction. The Hamiltonian is 
assumed to be invariant under color $SU(N)$. Since the adjoint of
$(\phi^A)_\beta^\alpha$ has the color transformation properties of
$(\phi^A)_\alpha^\beta$ it is convenient to define
the creation operator $({\bar\phi^A})_\alpha^\beta
=((\phi^A)_\beta^\alpha)^\dagger$.

Lower case Latin letters
$a_k=1,\ldots, s$. are spin indices, with the number of spin indices 
$n=0,1,\ldots,s$. Each operator is completely antisymmetric under 
permutations of the $a$'s. There are $2^s$ distinct sets of these indices. 
The Hamiltonian is assumed to be invariant under $O(s)$ rotations of these
indices. For generic $s$ these are vector indices, but for some special
values of $s$, they may alternatively be taken as spinor indices.
Whether or not the string bit system leads to a Poincar\'e invariant
string theory we specify that $\phi$ satisfies commutation relations
when $n$ is even and anticommutation relations when $n$ is odd.
For $s>0$ there are an equal number of bosonic and fermionic
annihilation operators.
With this assignment of statistics, the spin statistics theorem implies
that only values of $s$ that admit spinor representations can
lead to the emergence of Poincar\'e invariance. In the
case of principal interest for string theory, $s=8$ admits the
SO(8) spinors comprising the physical Majorana-Weyl degrees
of freedom required of the $D=10$ superstring.
 In the case of the protostring $s=24$
can be taken as 3 $SO(8)$ Majorana spinors.

Finally, the capital Latin superscript $A=1,\ldots, N_f$ is
a label for internal
(non spin or ``flavor'') degrees of freedom. For the purposes of
this article we
make no assumptions about their structure. In \cite{thornspace}
it was shown how to choose their dynamics so as to generate a
Heisenberg spin system on a chain of bits, which simulates
noncompact transverse coordinates. Such a choice would presumably be
important for precisely inducing the superstring, but in this article
we leave their structure arbitrary. We have borrowed the terms color and flavor
from the standard model, because their roles in their respective
models are similar: String bits will be permanently confined in
color singlet string as quarks and gluons are in color singlet  hadrons.
And flavor distinguishes otherwise identical string bits as flavor
distinguishes otherwise identical quarks and leptons.

In string bit models, supersymmetry plays a fundamental role in the 
dynamics of string formation. Consider a closed chain of $M$ bits in the
limit of very large $M$. Stringy behavior can ensue if there are
excitation energies above the ground state of order $O(1/M)$.
This occurs for example if there are vibrational modes and also for
ordinary spin waves arising from fluctuations between different
spin states of each bit with the same statistics. In both cases
the large $M$ behavior of the
ground energy has the behavior $E(M)\sim\alpha M+\gamma/M$ with
$\gamma<0$. In our string bit models, at order $1/N$,
a closed chain can split into two
similar but smaller closed chains such that $M=M_1+M_2$. Then
$\gamma<0$ would mean that the longer chain is unstable to decay into
smaller chains (as long as the daughter chains are long enough).
This is ultimately the reason
the bosonic closed string as well as the Neveu-Schwarz (NS,NS)
closed string have tachyonic ground states.

If, however, there are waves
arising from fluctuations between bosonic and fermionic states of each string
bit, the sign of $\gamma$ reverses and the single string is stable: it
requires added energy to transform into two smaller chains. This is what
happens in the Green-Schwarz lightcone superstring in which the fermionic
worldsheet field $S$ is a rotational spinor in target space i.e. it creates
a target space fermionic state from a bosonic one and {\it vice versa}.
But for such $1/M$ excitations to be possible each bit must possess
a fermionic and
a bosonic state of the same energy: there must be at least one
Grassmann odd operator
$Q$ which commutes with the Hamiltonian. This is all that is
meant by supersymmetry in the following.
It is a lot less supersymmetry than the full supersymmetry of
an emergent superstring, which among other things would require another
Grassmann odd operator $R$ such that $H=R^2$. This larger supersymmetry
is not necessary for stable string formation, but rather is
crucial for the Poincar\'e invariance of any emergent string theory\footnote{
It is not obvious how much of the symmetry one finds in the emergent
string theory should or even can be present in the string bit Hamiltonian.
After all the latter makes no reference to space at all, to say nothing of
Super-Poincar\'e invariance.}.

In this article we explore ways to construct supersymmetric
string bit Hamiltonians in the presence of internal non-spin
degrees of freedom represented by the capital superscript in (\ref{bitann}).
In Section 2 we recall some essential features of the string bit formalism
that we need for this work. The construction of supersymmetric
polynomials quadratic and quartic in bit creation and annihilation
operators is developed in Section 3. Finally, in Section 4, we
conclude by applying the tools developed in Section 3 to build
quartic terms, which modify the Hamiltonian of \cite{thornspace} to supply
the necessary prefactors to produce superstring scattering amplitudes.

\section{Review of the string bit formalism}
The string bit creation and annihilation
 operators described in the introduction satisfy the
(anti)-commutation relations
\bea
{}[(\phi^A_{a_1\cdots a_m})_\alpha^{\ \beta},({\bar\phi}^B_{b_1\cdots b_n})_\gamma^{\ \delta}]_\pm&\equiv&(\phi^A_{a_1\cdots a_m})_\alpha^{\ \beta}
({\bar\phi}^B_{b_1\cdots b_n})_\gamma^{\ \delta}-(-)^{mn}({\bar\phi}^B_{b_1\cdots b_n})_\gamma^{\ \delta}(\phi^A_{a_1\cdots a_m})_\alpha^{\ \beta}\nonumber\\
&=&\delta_{AB}\delta_{mn}\delta_{\alpha}^{\ \delta}\delta_{\gamma}^{\ \beta}\sum_P(-)^P\delta_{a_1b_{P_1}}\cdots\delta_{a_nb_{P_n}},
\label{crs}
\eea
which incorporate the fact that ${\bar\phi}$ creates a boson if $n$ is
even and a fermion if $n$ is odd. The sum over $P$ is over all permutations
of $1,2,\ldots,n$.

In \cite{thornspace} we studied a Hamiltonian with a very specific
implementation of the dynamics of these internal ``flavor'' degrees of freedom.
We wrote $H=H_F+H_S$ with $H_F$ diagonal in spin,
but non-trivial in flavor, and {\it vice versa} for $H_S$
\bea
H_F&=&\frac{2}{N}\sum_{n=0}^s\sum_{k=0}^s\frac{1}{n!k!}
\Tr{\bar\phi}^A_{a_1\cdots a_n}{\bar\phi}^B_{b_1\cdots b_k}
{\phi}^C_{b_1\cdots b_k}{\phi}^D_{a_1\cdots a_n}T_{ABCD}
\eea
and for $H_S$:
\bea
H_S&=& H_1+H_2+H_3+H_4+H_5,
\label{sham}
\eea
where the $H_i$ are:
\bea
H_1&=&\frac{2}{N}\sum_{n=0}^s\sum_{k=0}^s\frac{s-2n}{n!k!}
\Tr{\bar\phi}^A_{a_1\cdots a_n}{\bar\phi}^B_{b_1\cdots b_k}
{\phi}^B_{b_1\cdots b_k}{\phi}^A_{a_1\cdots a_n}
\eea
\bea
H_2&=&\frac{2}{N}\sum_{n=0}^{s-1}\sum_{k=0}^{s-1}\frac{(-)^k}{n!k!}
\Tr{\bar\phi}^A_{a_1\cdots a_n}{\bar\phi}^B_{bb_1\cdots b_k}
{\phi}^B_{b_1\cdots b_k}{\phi}^A_{ba_1\cdots a_n}
\eea
\bea
H_3&=&\frac{2}{N}\sum_{n=0}^{s-1}\sum_{k=0}^{s-1}\frac{(-)^k}{n!k!}
\Tr{\bar\phi}^A_{ba_1\cdots a_n}{\bar\phi}^B_{b_1\cdots b_k}
{\phi}^B_{bb_1\cdots b_k}{\phi}^A_{a_1\cdots a_n}
\eea
\bea
H_4&=&\frac{2i}{N}\sum_{n=0}^{s-1}\sum_{k=0}^{s-1}\frac{(-)^k}{n!k!}
\Tr{\bar\phi}^A_{a_1\cdots a_n}{\bar\phi}^B_{b_1\cdots b_k}
{\phi}^B_{bb_1\cdots b_k}{\phi}^A_{ba_1\cdots a_n}
\eea
\bea
H_5&=&-\frac{2i}{N}\sum_{n=0}^{s-1}\sum_{k=0}^{s-1}\frac{(-)^k}{n!k!}
\Tr{\bar\phi}^A_{ba_1\cdots a_n}{\bar\phi}^B_{bb_1\cdots b_k}
{\phi}^B_{b_1\cdots b_k}{\phi}^A_{a_1\cdots a_n}.
\eea
Torres \cite{torres} has shown that both $H_F$ and $H_S$ 
commute with the supersymmetry operators
\bea
Q^a&=&\sum_{n=0}^{s-1}\frac{(-)^n}{n!}\Tr\left[{\gamma}
{\bar\phi}^A_{a_1\cdots a_n}\phi^A_{aa_1\cdots a_n}+{\gamma^*}
{\bar\phi}^A_{aa_1\cdots a_n}\phi^A_{a_1\cdots a_n}\right],\qquad \gamma
\equiv e^{i\pi/4}\label{susyQ}\\
\{Q^a,Q^b\}&=&2MN_f\delta_{ab}\\
M&=&\sum_{n=0}^{s}\frac{1}{n!}\Tr
{\bar\phi}^A_{a_1\cdots a_n}\phi^A_{a_1\cdots a_n},
\label{bitnumber}
\eea
where $M$ is the bit number operator, $N_F$ is the number of flavors,
and all repeated indices are summed.
We have introduced the complex number $\gamma$ to
reduce clutter in subsequent equations. The vanishing commutator of $Q^a$
with $H$ guarantees equal numbers of bosonic and fermionic eigenstates,
at each energy level with bit number greater than 0.
In this article we will extend this conclusion to
more general Hamiltonians.

Using the commutation relations (\ref{crs}), it is straightforward 
to obtain the action of the $H_i$ on single trace states. We recall
from \cite{bergmantsubit,sunthorn,thornspace} the definition of superfields
\bea
{\bar\psi}^A(\theta)&=&\sum_{k=0}^s\frac{1}{k!}{\bar\phi}^A_{a_1\cdots a_k}
\theta^{a_1}\cdots\theta^{a_k}
\eea
where $\theta^a$ are Grassmann variables, and using the notation
\bea
T(A_1,\theta_1;\cdots;A_M,\theta_M)&=&\Tr{\bar\psi}^{A_1}(\theta_1),
\cdots,{\bar\psi}^{A_M}(\theta_M),
\eea
we find (suppressing flavor indices)
\bea
H_1T(\theta_1,\cdots,\theta_M)\ket{0}&=&2\sum_{k=1}^M\left(s-2\theta_k^a
\frac{d}{d\theta_k^a}\right)
T(\theta_1,\cdots,\theta_M)\ket{0}\nonumber\\
&&\hskip-.25in +\frac{2}{N}\sum_{k=1}^M\left(s-2\theta_k^a
\frac{d}{d\theta_k^a}\right)\sum_{l\neq k,k+1}T(\theta_l\cdots\theta_k)
T(\theta_{k+1}\cdots\theta_{l-1})\ket{0}
\eea
\bea
H_2T(\theta_1,\cdots,\theta_M)\ket{0}&=&2\sum_{k=1}^M
\theta_k^a\frac{d}{d\theta_{k+1}^a}T(\theta_1,\cdots,\theta_M)\ket{0}\nonumber\\
&&\hskip-.25in +\frac{2}{N}\sum_{k=1}^M\sum_{l\neq k,k+1}
\theta_k^a\frac{d}{d\theta_{l}^a}T(\theta_l\cdots\theta_k)
T(\theta_{k+1}\cdots\theta_{l-1})\ket{0}
\eea
\bea
H_3T(\theta_1,\cdots,\theta_M)\ket{0}&=&2\sum_{k=1}^M
\theta_{k+1}^a\frac{d}{d\theta_{k}^a}T(\theta_1,\cdots,\theta_M)\ket{0}\nonumber\\
&&\hskip-.25in +\frac{2}{N}\sum_{k=1}^M\sum_{l\neq k,k+1}
\theta_l^a\frac{d}{d\theta_{k}^a}T(\theta_l\cdots\theta_k)
T(\theta_{k+1}\cdots\theta_{l-1})\ket{0}
\eea
\bea
H_4T(\theta_1,\cdots,\theta_M)\ket{0}&=&-2i\sum_{k=1}^M
\theta_k^a\theta^a_{k+1}T(\theta_1,\cdots,\theta_M)\ket{0}\nonumber\\
&&\hskip-.25in -\frac{2i}{N}\sum_{k=1}^M\sum_{l\neq k,k+1}
\theta_k^a{\theta_{l}^a}T(\theta_l\cdots\theta_k)
T(\theta_{k+1}\cdots\theta_{l-1})\ket{0}
\\
H_5T(\theta_1,\cdots,\theta_M)\ket{0}&=&-2i\sum_{k=1}^M
\frac{d}{d\theta_k^a}\frac{d}{d\theta^a_{k+1}}T(\theta_1,\cdots,\theta_M)\ket{0}
\nonumber\\
&&\hskip-.25in -\frac{2i}{N}\sum_{k=1}^M\sum_{l\neq k,k+1}
\frac{d}{d\theta_k^a}\frac{d}{d\theta_{l}^a}T(\theta_l\cdots\theta_k)
T(\theta_{k+1}\cdots\theta_{l-1})\ket{0}.\eea
We note that the differential operators are applied to nearest neighbors on the
same trace when they involve two distinct Grassmann variables.

The action of the $H_i$ on multi-trace states takes two forms. 
When both annihilation operators contract on the same trace,
 the action can be read off from the
preceding formulas. When they act on different traces the action is to
fuse them into a single trace as follows
\bea
H_1T({\theta_1\cdots\theta_K})
T({\eta_{1}\cdots\eta_L})\ket{0}_{\rm Fusion}&=&
\nonumber\\
&&\hskip-2.5in +\frac{2}{N}\sum_{k=1}^K\sum_{l=1}^L \left(s-2\theta_k^a
\frac{d}{d\theta_k^a}\right)T({\theta_{k+1}
\cdots\theta_{k}}\eta_l\cdots\eta_{l-1})\ket{0}
\nonumber\\
&&\hskip-2.5in +\frac{2}{N}\sum_{k=1}^K\sum_{l=1}^L  \left(s-2\eta_l^a
\frac{d}{d\eta_l^a}\right)T(\theta_k\cdots\theta_{k-1}{\eta_{l+1}
\cdots\eta_{l}})\ket{0}\\
H_2T({\theta_1\cdots\theta_K})
T({\eta_{1}\cdots\eta_L})\ket{0}_{\rm Fusion}&=&
\nonumber\\
&&\hskip-2.5in +\frac{2}{N}\sum_{k=1}^K\sum_{l=1}^L \theta_k^a\frac{d}{d\eta_l}T({\theta_{k+1}
\cdots\theta_{k}}\eta_l\cdots\eta_{l-1})\ket{0}
\nonumber\\
&&\hskip-2.5in +\frac{2}{N}\sum_{k=1}^K\sum_{l=1}^L  
\eta_l^a\frac{d}{d\theta_k}T(\theta_k\cdots\theta_{k-1}{\eta_{l+1}
\cdots\eta_{l}})\ket{0}.\label{fuse1}
\eea
with similar transcriptions for the other $H_i$. Here the sequence
of labels $k\cdots (k-1)$ means $k\cdots K,1\cdots (k-1)$
and similarly for $l\cdots (l-1)$.
  In each case the differential
operators have the same structure as the fission terms, but the states on the
right are a suitable pair of single trace states. And when there are two 
distinct Grassmann operators they act on nearest neighbors on the large trace.

\section{Supersymmetry with Flavor}
The bit annihilation operator with flavor has, in addition to spinor indices,
another ``flavor''  index not associated with spin, so we write
$\phi^A_{a_1\cdots a_k}$ and we take the supercharge to be given
by (\ref{susyQ}).
Note that by construction $Q^a$ is hermitian. Our goal is to construct 
Hamiltonians which commute with $Q^a$ and with $M$. To this end we work out the
(anti)commutators ($[A,B]_\pm\equiv AB\pm BA$, with the + sign chosen
only when both $A$ and $B$ are Grassmann odd): 
\bea
{}[Q^a,{\bar\phi}^A_{b_1\cdots b_k}]_\pm&=&(-)^k{\gamma^*}{\bar\phi}^A_{ab_1\cdots b_k}
+{\gamma}\sum_{l=1}^k(-)^{l+k}\delta_{ab_l}{\bar\phi}^A_{b_1\cdots{\hat b}_l\cdots b_k}\\
{}[Q^a,{\phi}^A_{b_1\cdots b_k}]_\pm&=&-{\gamma}{\phi}^A_{ab_1\cdots b_k}
+{\gamma^*}\sum_{l=1}^k(-)^{l+1}\delta_{ab_l}{\phi}^A_{b_1\cdots{\hat b}_l\cdots b_k}
\eea
where a hat over an index means that that index is deleted.
The second line follows from the hermitian conjugate of the first line.
\subsection{Supersymmetry transform of some spinor-tensor bilinears}
For two matrix operators $A,B$ introduce the notation $A\otimes B$ to denote
the Hilbert space operator products of independent matrix elements of $A$ and
$B$. This notation allows us to suppress color indices in the derivations to follow. When explicit indices are needed we will use the following convention
\bea
(A\otimes B)_{\alpha\gamma}^{\beta\delta}\equiv A_\alpha^\beta B_\gamma^\delta.
\eea
The hermitian conjugate of these bilinears is given by
\bea
\left[(A\otimes B)_{\alpha\gamma}^{\beta\delta}\right]^\dagger=
(B_\gamma^\delta)^\dagger (A_\alpha^\beta)^\dagger
={\bar B}_\delta^\gamma{\bar A}_\beta^\alpha=({\bar B}\otimes{\bar A})_{\delta\beta}^{\gamma\alpha}
\eea
In addition to taking the trace of $A$ or $B$, one can form a matrix operator
from $A\otimes B$ in two ways, which we distinguish by a right arrow over the operator if we contract $\beta$ with $\gamma$ and
a left arrow if we contract $\alpha$ with $\delta$. Thus if $C=A\otimes B$
we have
\bea
{\overrightarrow C}_\alpha^\beta&\equiv& C_{\alpha\gamma}^{\gamma\beta}
=A_\alpha^\gamma B_\gamma^\beta=(AB)_\alpha^\beta\\
{\overleftarrow C}_\alpha^\beta&\equiv& C_{\gamma\alpha}^{\beta\gamma}
=A_\gamma^\beta B_\alpha^\gamma
\eea
We shall usually choose the first way which corresponds to usual matrix multiplication $AB$. The second way would correspond to $BA$ only if the
operator matrix elements all commute.
Then we can define several bilinear operators with simple supersymmetry transformation rules. Two bosonic bilinears are
\bea
\rho_{AB}&=&\sum_{n=0}^s\frac{1}{n!} {\bar\phi}^A_{b_1\cdots b_n}\otimes{\phi}^B_{b_1\cdots b_n}\\
\rho^{\prime}_{AB}&=&\sum_{n=0}^s\frac{n}{n!} {\bar\phi}^A_{b_1\cdots b_n}\otimes{\phi}^B_{b_1\cdots b_n}
\eea 
One must keep in mind that these bilinears have 4 suppressed color indices,
a pair for each of the factor operators. Since $Q^a$ is a color 
singlet one body operator, its commutator with any bilinear 
leaves these color indices and the ordering of operators undisturbed.
Then we calculate the commutators
\bea
&&\hskip-.75in{}[Q^a,\rho_{AB}]=
\sum_{n=0}^s\frac{1}{n!}\left([Q^a,{\bar\phi}^A_{b_1\cdots b_n}]_\pm\otimes{\phi}^B_{b_1\cdots b_n}+
(-)^n{\bar\phi}^A_{b_1\cdots b_n}\otimes
[Q^a,{\phi}^B_{b_1\cdots b_n}]_\pm\right)\nonumber\\
&=&\sum_{n=0}^s\frac{1}{n!}\left((-)^n{\gamma^*}{\bar\phi}^A_{ab_1\cdots b_n}
+{\gamma}\sum_{l=1}^n(-)^{l+n}\delta_{ab_l}{\bar\phi}^A_{b_1\cdots{\hat b}_l\cdots b_n}\right)\otimes{\phi}^B_{b_1\cdots b_n}\nonumber\\
&&+\sum_{n=0}^s\frac{1}{n!}(-)^n{\bar\phi}^A_{b_1\cdots b_n}\otimes
\left(-{\gamma}{\phi}^B_{ab_1\cdots b_n}
+{\gamma^*}\sum_{l=1}^n(-)^{l+1}\delta_{ab_l}{\phi}^B_{b_1\cdots{\hat b}_l\cdots b_n}\right)\nonumber\\
&=&\sum_{n=0}^s\left(\frac{(-)^n}{n!}{\gamma^*}{\bar\phi}^A_{ab_1\cdots b_n}
\otimes{\phi}^B_{b_1\cdots b_n}+\frac{(-)^{n-1}}{(n-1)!}{\gamma}{\bar\phi}^A_{b_1\cdots b_{n-1}}\otimes{\phi}^B_{ab_1\cdots b_{n-1}}\right)\nonumber\\
&&+\sum_{n=0}^s
\left(-\frac{(-)^n}{n!}{\gamma}
{\bar\phi}^A_{b_1\cdots b_n}\otimes{\phi}^B_{ab_1\cdots b_n}
-\frac{(-)^{n-1}}{(n-1)!}{\gamma^*}
{\bar\phi}^A_{ab_1\cdots b_{n-1}}\otimes{\phi}^B_{b_1\cdots b_{n-1}}\right)
\nonumber\\
&=&0
\eea
after shifting the sum index of the second terms in each set of parentheses
$n\to n+1$.

For $\rho^\prime$, we have
\bea
&&\hskip-.75in{}[Q^a,\rho_{AB}^\prime]=
\sum_{n=0}^s\left(\frac{(-)^n}{(n-1)!}{\gamma^*}{\bar\phi}^A_{ab_1\cdots b_n}
\otimes{\phi}^B_{b_1\cdots b_n}+\frac{n(-)^{n-1}}{(n-1)!}{\gamma}{\bar\phi}^A_{b_1
\cdots b_{n-1}}\otimes{\phi}^B_{ab_1\cdots b_{n-1}}\right)\nonumber\\
&&+\sum_{n=0}^s
\left(-\frac{(-)^n}{(n-1)!}{\gamma}
{\bar\phi}^A_{b_1\cdots b_n}\otimes{\phi}^B_{ab_1\cdots b_n}
-\frac{n(-)^{n-1}}{(n-1)!}{\gamma^*}
{\bar\phi}^A_{ab_1\cdots b_{n-1}}\otimes{\phi}^B_{b_1\cdots b_{n-1}}\right)
\nonumber\\
&=&\sum_{n=0}^s\frac{(-)^{n}}{n!}\left({\gamma}{\bar\phi}^A_{b_1
\cdots b_{n}}\otimes{\phi}^B_{ab_1\cdots b_{n}}-{\gamma^*}
{\bar\phi}^A_{ab_1\cdots b_{n}}\otimes{\phi}^B_{b_1\cdots b_{n}}\right)
\nonumber\\
&\equiv&{\gamma}\eta^a_{AB}-{\gamma^*}{\bar\eta}^a_{AB}
\eea
The next to last line is obtained by writing $n=n-1+1$ 
in the second terms inside parentheses on the previous two lines 
and then recognizing that the ``n-1'' parts
cancel against the first terms, 
leaving the last line with the $n$ summation index shifted by 1.

We can similarly consider the supersymmetry properties of the Grassmann odd
bilinears $\eta,{\bar\eta}$, defined in the previous equation
\bea
\eta_{AB}^a&=&\sum_{n=0}^{s-1}\frac{(-)^n}{n!}
{\bar\phi}^A_{a_1\cdots a_n}\otimes\phi^B_{aa_1\cdots a_n}
\eea
Forming ${\gamma}\eta^a_{AA}+{\gamma^*}\eta^{a\dagger}_{AA}$
saturating the color indices as the trace of a matrix product, and 
summing over all flavors $A$ gives the supercharge $Q^a$.
Then we calculate
\bea
\{Q^a,\eta_{AB}^b\}&=&\sum_{n=0}^{s-1}\frac{(-)^n}{n!}
\left((-)^n{\gamma^*}{\bar\phi}^A_{ab_1\cdots b_n}
+{\gamma}\sum_{l=1}^n(-)^{l+n}\delta_{ab_l}
{\bar\phi}^A_{b_1\cdots{\hat b}_l\cdots b_n}\right)
\otimes\phi^B_{bb_1\cdots b_n}\nonumber\\
&&+\sum_{n=0}^s\frac{1}{n!}{\bar\phi}^A_{b_1\cdots b_n}\otimes
\left(-{\gamma}{\phi}^B_{abb_1\cdots b_n}   
+{\gamma^*}\sum_{l=1}^n(-)^{l}\delta_{ab_l}
{\phi}^B_{bb_1\cdots{\hat b}_l\cdots b_n}\right)\nonumber\\
&&+\delta_{ab}{\gamma^*}\sum_{n=0}^s\frac{1}{n!}{\bar\phi}^A_{b_1\cdots b_n}
\otimes{\phi}^B_{b_1\cdots b_n}\nonumber\\
&=&\sum_{n=0}^{s-1}\frac{1}{n!}
\left({\gamma^*}{\bar\phi}^A_{ab_1\cdots b_n}\otimes\phi^B_{bb_1\cdots b_n}
+n{\gamma}
{\bar\phi}^A_{b_1\cdots b_{n-1}}\otimes\phi^B_{abb_1\cdots b_{n-1}}\right)
\nonumber\\
&&+\sum_{n=0}^s\frac{1}{n!}
\left(-{\gamma}{\bar\phi}^A_{b_1\cdots b_n}\otimes
{\phi}^B_{abb_1\cdots b_n}   
-n{\gamma^*}
{\bar\phi}^A_{b_1\cdots b_{n-1}}\otimes{\phi}^B_{bb_1\cdots b_{n-1}}\right)
\nonumber\\
&&+\delta_{ab}{\gamma^*}\sum_{n=0}^s\frac{1}{n!}{\bar\phi}^A_{b_1\cdots b_n}
\otimes{\phi}^B_{b_1\cdots b_n}\nonumber\\
&=&\delta_{ab}{\gamma^*}\sum_{n=0}^s\frac{1}{n!}{\bar\phi}^A_{b_1\cdots b_n}
\otimes{\phi}^B_{b_1\cdots b_n}
\eea
By taking the Hermitian conjugate, we learn that
\bea
\{Q^a,\eta^{b\dagger}_{BA}\}&=&\delta_{ab}{\gamma}
\sum_{n=0}^s\frac{1}{n!}{\bar\phi}^A_{b_1\cdots b_n}
\otimes{\phi}^B_{b_1\cdots b_n}
\eea
and then it immediately follows that $\{Q^a,Q^b\}=2MN_f\delta_{ab}$ where
$M=\sum_n\frac{1}{n!}\Tr{\bar\phi}^A_{b_1\cdots b_n}{\phi}^A_{b_1\cdots b_n}$
is the bit number operator. Another immediate corollary is
\bea
\{Q^a,\gamma\eta^b_{AB}-\gamma^*\eta^{b\dagger}_{BA}\}&=&0
\eea

This process of building bilinears can be extended to higher rank
spinor-tensors.
Consider the  bilinears
\bea
\Omega^{AB,p}_{a_1\cdots a_l}&=&\sum_k\frac{(-)^{kl}}{k!}{\bar\phi}^A_{a_1\cdots a_pb_1\cdots b_k}\otimes\phi^B_{a_{p+1}\cdots a_lb_1\cdots b_k}
\eea
which includes as special cases the bilinears $\rho$, $\eta^a$ and $
{\bar\eta}^a$ discussed above. In an appendix, we show that the
(anti)commutators of $Q^c$ with these bilinears are
\bea
{}[Q^c,\Omega^{AB,p}_{a_1\cdots a_l}]_{\pm}&=&\gamma\sum_{m=1}^p(-)^{m+p}
\delta_{ca_m}\Omega^{AB,p-1}_{a_1\cdots {\hat a}_m\cdots a_l}
-\gamma^*\sum_{m=p+1}^l(-)^{m}
\delta_{ca_m}\Omega^{AB.p}_{a_1\cdots {\hat a}_m\cdots a_l}
\label{Qanticom}
\eea
where $\gamma= e^{i\pi/4}$.
\subsection{Supersymmetric Bilinears}
Among the bilinears constructed in the last subsection we have noted that
$\rho^{AB}$ and the linear combination
$\gamma\eta^b_{AB}-\gamma^*\eta^{b\dagger}_{BA}$
are supersymmetric i.e. they commute and anticommute, respectively,
with $Q^c$. In fact they are just the first two
of a sequence of such bilinears, we can form from linear
combinations of the $\Omega^{AB,p}_{a_1\cdots a_l}$. In the following
we suppress the flavor labels $A,B$ since they are simply spectators.

For a given $l$, let $E^{a_1\cdots a_l}$ be a completely antisymmetric spinor-tensor of rank $l$. For $l=s$ the unique choice is the Levi-Cevita tensor. But
for $l<s$ there are multiple choices. Then we calculate
\bea
{}[Q^c,E^{a_1\cdots a_l}\Omega^p_{a_1\cdots a_l}]_{\pm}
&=&\gamma\sum_{m=1}^p(-)^{m+p}E^{a_1\cdots c\cdots a_l}\Omega^{p-1}_{a_1\cdots
  {\hat a}_m\cdots a_l}-\gamma^*\sum_{m=p+1}^l(-)^{m}E^{a_1\cdots c\cdots a_l}
\Omega^{p}_{a_1\cdots
  {\hat a}_m\cdots a_l}\nonumber\\
&=&\gamma(-)^{p+1}pE^{ca_1\cdots a_{l-1}}\Omega^{p-1}_{a_1\cdots
   a_{l-1}}+\gamma^*(l-p)E^{ca_1\cdots a_{l-1}}
 \Omega^{p}_{a_1\cdots a_{l-1}}\eea
 In the first line the index $c$ occupies the $m$th location in the superscript of $E$. Then the last line is obtained by moving $c$ to the first location
 and relabeling the summation indices $a_1\cdots
 {\hat a}_m\cdots a_l\to a_1\cdots a_{l-1}$.

 Now make the {\it ansatz}
 \bea
 \Omega^{AB}_{E_l}=\sum_{p=0}^lc_pE^{a_1\cdots a_l}\Omega^{AB,p}_{a_1\cdots a_l}
 \eea
 and require, still suppressing flavor indices,
 \bea
 0&=&[Q^c,\Omega_E]_\pm=\sum_{p=0}^lc_p\left[\gamma(-)^{p+1}pE^{ca_1\cdots a_{l-1}}\Omega^{p-1}_{a_1\cdots
   a_{l-1}}+\gamma^*(l-p)E^{ca_1\cdots a_{l-1}}
 \Omega^{p}_{a_1\cdots a_{l-1}}\right]\nonumber\\
&=&\sum_{p=0}^{l-1}\left[c_{p+1}\gamma(-)^p(p+1)+c_p\gamma^*(l-p)\right]
E^{ca_1\cdots a_{l-1}}
\Omega^{p}_{a_1\cdots a_{l-1}}
\eea
which implies the recursion relation
\bea
c_{p+1}=-c_p\frac{(-)^p\gamma^*(l-p)}{\gamma(p+1)}
\eea
with the solution, setting $c_0=\gamma^l$,
\bea
c_n=(-)^{n(n+1)/2}\gamma^{*n}\gamma^{l-n}{l\choose n}.
\eea
\subsection{Supersymmetric Quartics}
We now turn to the search for candidates for a
supersymmetric string bit Hamiltonian, under the working hypothesis that
$H$ will be quartic in the string bit creation and annihilation operators.
The simplest possibilities are to take the products of pairs of
the supersymmetric bilinears given in the previous section:
\bea
:\Omega^{AB}_{E_l}\otimes\Omega^{CD}_{F_{l^\prime}}:
\eea
is supersymmetric by construction. Except for $l,l^\prime=0$ or $=s$,
$E$ and $F$ break the $O(s)$ symmetry, which could be restored either by
suitable averages over $E,F$, or by adding new degrees of freedom. such
as worldsheet coordinate fields, to compensate. 
Here, as before,  the
$\otimes$ symbol signifies that none of the suppressed color indices
are contracted. Of course, since we require that the Hamiltonian be a color
singlet, all these indices must be contracted. For example, the term
in $H_1$ proportional to $s$ can be generalized to
\bea
\frac{2}{N}\sum_{n=0}^s\sum_{k=0}^s\frac{s}{n!k!}
\Tr{\bar\phi}^A_{a_1\cdots a_n}{\bar\phi}^B_{b_1\cdots b_k}
{\phi}^C_{b_1\cdots b_k}{\phi}^D_{a_1\cdots a_n}
\eea
which can be obtained from $:\rho_{AD}\otimes\rho_{BC}:$ by the index contraction
$:(\rho_{AD})_{\alpha\delta}^{\beta\alpha}(\rho_{BG})_{\beta\gamma}^{\gamma\delta}:$.
It commutes with $Q^a$ since the latter commutes with $\rho_{AB}$. But any other
contraction scheme will also preserve supersymmetry, and any such term can e
included in modifying the Hamiltonian keeping supersymmetry. The normal ordering
does not interfere with the product rule for commutators, because $Q^a$ is
a one body operator.

Next we implement a similar generalization of $H_{2-5}$.
\bea
H_2&\to&\frac{2}{N}\sum_{n=0}^{s-1}\sum_{k=0}^{s-1}\frac{(-)^k}{n!k!}
\Tr{\bar\phi}^A_{a_1\cdots a_n}{\bar\phi}^B_{bb_1\cdots b_k}
{\phi}^C_{b_1\cdots b_k}{\phi}^D_{ba_1\cdots a_n}\\
H_3&\to&\frac{2}{N}\sum_{n=0}^{s-1}\sum_{k=0}^{s-1}\frac{(-)^k}{n!k!}
\Tr{\bar\phi}^A_{ba_1\cdots a_n}{\bar\phi}^B_{b_1\cdots b_k}
{\phi}^C_{bb_1\cdots b_k}{\phi}^D_{a_1\cdots a_n}\\
H_4&\to&\frac{2i}{N}\sum_{n=0}^{s-1}\sum_{k=0}^{s-1}\frac{(-)^k}{n!k!}
\Tr{\bar\phi}^A_{a_1\cdots a_n}{\bar\phi}^B_{b_1\cdots b_k}
{\phi}^C_{bb_1\cdots b_k}{\phi}^D_{ba_1\cdots a_n}\\
H_5&\to&-\frac{2i}{N}\sum_{n=0}^{s-1}\sum_{k=0}^{s-1}\frac{(-)^k}{n!k!}
\Tr{\bar\phi}^A_{ba_1\cdots a_n}{\bar\phi}^B_{bb_1\cdots b_k}
{\phi}^C_{b_1\cdots b_k}{\phi}^D_{a_1\cdots a_n}.
\eea
The right sides can be obtained, respectively from
\bea
-\frac{2}{N}:{\eta}^b_{AD}\otimes{\bar\eta}^b_{BC}:,\quad
\frac{2}{N}:{\bar\eta}^b_{AD}\otimes\eta^b_{BC}:,\quad
-\frac{2i}{N}:{\eta}^b_{AD}\otimes\eta^b_{BC}:,\quad
-\frac{2i}{N}:{\bar\eta}^b_{AD}\otimes{\bar\eta}^b_{BC}:
\eea
Since the factor operators are not supersymmetric, the $H_i$ are
not separately supersymmetric, but we shall see shortly that the
sum of the modified terms is supersymmetric.

The commutator of $Q^a$ with each of these expressions is given respectively by
\bea
&&\frac{2}{N}\left({\gamma}:\eta^a_{AD}\otimes\rho_{BC}:
-{\gamma^*}:\rho_{AD}\otimes{\bar\eta}^a_{BC}:\right)\\
&&\frac{2}{N}\left({\gamma}:\rho_{AD}\otimes\eta^a_{BC}:
-{\gamma^*}:{\bar\eta}^a_{AD}\otimes\rho_{BC}:\right)\\
&&\frac{2}{N}\left({\gamma}:\eta^a_{AD}\otimes\rho_{BC}:
-{\gamma}:\rho_{AD}\otimes\eta^a_{BC}:\right)\\
&&\frac{2}{N}\left({\gamma^*}:\rho_{AD}\otimes{\bar\eta}^a_{BC}:
-{\gamma^*}:{\bar\eta}^a_{AD}\otimes\rho_{BC}:\right)
\eea 
Notice that the sum of these four expressions is
\bea
\frac{4}{N}\left({\gamma}:\eta^a_{AD}\otimes\rho_{BC}:
-{\gamma^*}:{\bar\eta}^a_{AD}\otimes\rho_{BC}:\right)
\eea 
which can be cancelled by 
\bea
{}\frac{4}{N}[Q^a,:\rho^\prime_{AD}\otimes\rho_{BC}:]&=&
\frac{4}{N}:\left({\gamma}\eta^a_{AD}-{\gamma^*}{\bar\eta}^a_{AB}
\right)\otimes\rho_{BC}:
\eea
Similarly, the sum of the first two lines minus the sum of the last two lines
is
\bea
\frac{4}{N}\left({\gamma}:\rho^a_{AD}\otimes\eta^a_{BC}:
-{\gamma^*}:\rho_{AD}\otimes{\bar\eta}^a_{BC}:\right)
\eea
which can be cancelled by
\bea
{}\frac{4}{N}[Q^a,:\rho_{AD}\otimes\rho^\prime_{BC}:]&=&
\frac{4}{N}:\rho_{AD}\otimes\left({\gamma}\eta^a_{BC}-{\gamma^*}{\bar\eta}^a_{BC}
\right):
\eea
To summarize, in addition to the numerous supersymmetric quartics that can be
written as products of supersymmetric bilinears,
we have constructed two  supersymmetric structures, as bilinears
in the nonsupersymmetric bilinears $\eta^a, {\bar\eta}^a$, which enter
the $H_S$ part of the original string bit Hamiltonian. We have not tried
to find other such structures which are products of bilinears of tensors
of higher rank, though we expect many more should exist.
The structures entering $H_S$ are
\bea
&&\frac{2s}{N}:\rho^{AD}\otimes\rho^{BC}:\\
&&\frac{2}{N}\left(-:{\eta}^b_{AD}\otimes{\bar\eta}^b_{BC}:+
:{\bar\eta}^b_{AD}\otimes\eta^b_{BC}:
-i:{\eta}^b_{AD}\otimes\eta^b_{BC}
-i:{\bar\eta}^b_{AD}\otimes{\bar\eta}^b_{BC}:\right.\nonumber\\
&&\hskip.25in \left. -:\rho^\prime_{AD}\otimes\rho_{BC}\right)\\
&&\frac{2}{N}\left(-:{\eta}^b_{AD}\otimes{\bar\eta}^b_{BC}:+
:{\bar\eta}^b_{AD}\otimes\eta^b_{BC}:
+i:{\eta}^b_{AD}\otimes\eta^b_{BC}
+i:{\bar\eta}^b_{AD}\otimes{\bar\eta}^b_{BC}:\right.\nonumber\\
&&\hskip.25in \left. -:\rho_{AD}\otimes\rho^\prime_{BC}\right)
\eea
where the first is a product of supersymmetric bilinears, and
the last two are (at least apparently) not of this type..
To build a supersymmetric Hamiltonian from these structures, simply 
multiply each of these forms by an independent $T_k^{ABCD}$ and sum 
over all flavors. All color indices need to be contracted, but
supersymmetry holds for all
contraction schemes.
\section{Applications to the IIB Superstring}
In \cite{thornspace} the connection of string bit models to
string formation has been established in the limit $N\to\infty$,
which corresponds to zero string coupling.
In that limit $H_S$ on multi-trace states acts independently on each trace
with only nearest neighbor interactions of the bits on the trace,
identical with those of a discretized lightcone quantized Green-Schwarz
Grassmann
world sheet field. Interactions between these discretized strings are
present in $H_S$  at order $1/N$, but they do not provide the operator prefactor
required for Poincar\' e invariance in the continuum limit.
To describe the IIB superstring, additional terms must be added to $H_S$
which supply these prefactors at order $1/N$, while not contributing
at $N=\infty$.

In this section we determine supersymmetric terms, involving the $\Omega^p$
discussed in the previous section, which can provide the necessary prefactors.
In addition
to the 8 pairs of Grassmann worldsheet fields $S^a(\sigma,\tau)$,
$ {\tilde S}^a(\sigma,\tau)$,
the superstring requires 8 bosonic transverse 
coordinates $x^k(\sigma,\tau)$. In \cite{thornspace} we argued that these
transverse coordinates might actually be merely effective fields,
which describe the low energy dynamics of long Heisenberg spin chains,
in which the spin
matrices act on internal degrees of freedom, which we
have called ``flavor''.
However for the purposes of this discussion
we take them as fundamental worldsheet fields rather than effective ones.
Green, Schwarz, and Brink \cite{greenschwarzbrink}
derived the vertex insertion required by
Poincar\' e invariance. It has the structure
\bea
{\tilde X}^j(\sigma_I,\tau_I) X^k(\sigma_I,\tau_I)V_{jk}(Y(\sigma_I,\tau_I))
\label{fullinsertion}
\eea
with $V_{jk}$ a linear combination of the five structures
\bea
&&\delta_{jk},\quad \gamma^{jk}_{ab}Y^aY^b,\quad t^{jk}_{abcd}Y^aY^bY^cY^d,
\quad \gamma^{jk}_{ab}\epsilon^{abcdefgh}Y^cY^dY^eY^fY^gY^h,\nonumber\\
&&\delta_{jk}\epsilon^{abcdefgh}Y^aY^bY^cY^dY^eY^fY^gY^h
\label{insertions}
\eea
In these expressions ${\tilde X}$, $X$ are left and right moving
derivatives of the transverse coordinate worldsheet fields $x^k(\sigma,\tau)$:
\bea
X({\tilde X})&\propto& \lim_{\sigma\to\sigma_I} \sqrt{\sigma_I-\sigma}[{\cal P}(\sigma)\mp x^\prime(\sigma)]
\eea
where ${\cal P}={\dot x}$ is the momentum density conjugate to
$x$ and where the limit is taken {\it at the end} of the calculation. Also the
point on the worldsheet where a single string separates into
two strings is marked by $\sigma_I, \tau_I$.
The square root
factor is necessary because the interaction point is singular, making the
insertion blow up exactly at $\sigma=\sigma_I$. This is a drawback of
trying to work with a continuous world sheet.
This drawback is not present in the
string bit approach because in those models the emergent worldsheet
is discretized and behaves as a continuous worldsheet only for low
energy excitations. In string bit models the $X$ or ${\tilde X}$
insertion is simply $p_k\pm (x_{k+1}-x_k)$. 
The $Y^a$ are linear combinations of the
Grassmann worldsheet fields.
\bea
Y&\propto& \lim_{\sigma\to\sigma_I} \sqrt{\sigma_I-\sigma}
[\theta(\sigma)+\delta/\delta\theta(\sigma)]. 
\label{yinsertcont}
\eea
Again, the limiting procedure is necessary because the worldsheet is
continuous. In this case it gives the misleading impression that the
different powers of $Y$ in the complete insertion have different
divergent behavior. But, as shown in \cite{thornprotobits}, because the
relation of a discretized local Grassmann variable such as $\theta_k$
to energy ladder operators $B_n$ is of the form
\bea
\theta_k&\sim& \frac{1}{\sqrt{M}}\sum_{n=0}^{M-1}e^{2\pi i nk/M}B_n,
\eea
the square root
divergence ($\sim \sqrt{M}$) indicated by (\ref{yinsertcont})
is precisely what is needed to leave 
a finite nonzero effect in the continuum limit  ($M\to\infty$),
when the insertion is applied
to energy eigenstates! In string bit models the $Y$ dependent factor of the
insertion will simply be sums of powers of a linear combination of
$\theta_k$ and $\delta/\delta\theta_k$, with no further regulation.

We now show how these structures can be generated by supersymmetric terms added
to the string bit Hamiltonian. We start by working out the commutator of
$\Omega^{AB,p}_{E_l}$ with the superstring bit creation operator
${\bar\psi}^C(\theta)$. 
\bea
{}[\Omega^{AB,p}_{E_l},{\bar\psi}^C(\theta)]
&=&\delta_{BC}E^{a_1\cdots a_l}\sum_k\frac{(-)^{kl}}{k!}
{\bar\phi}^A_{a_1\cdots a_pb_1\cdots b_k}\theta^{a_{p+1}}\cdots\theta^{a_{l}}
\theta^{b_{1}}\cdots \theta^{b_{k}}\nonumber\\
&=&\delta_{BC}E^{a_1\cdots a_l}(-)^{p(l-p)}\theta^{a_{p+1}}\cdots\theta^{a_{l}}
\sum_k\frac{(-)^{kp}}{k!}
{\bar\phi}^A_{a_1\cdots a_pb_1\cdots b_k}
\theta^{b_{1}}\cdots \theta^{b_{k}}
\eea
To express the right side in terms of ${\bar\psi}(\theta)$ we
calculate
\bea
\frac{d}{d\theta^{a_p}}\cdots\frac{d}{d\theta^{a_1}}{\bar\psi}^A(\theta)&=&
\sum_{k=0}^s\frac{(-)^{p(p+k)}}{k!}
{\bar\phi}^A_{a_1\cdots a_pb_1\cdots b_k}\theta^{b_1}\cdots\theta^{b_k}
\eea
The sum on the right is proportional to the sum appearing in the commutator,
so we can write
\bea
{}[\Omega^{AB,p}_{E_l},{\bar\psi}^C(\theta)]
&=&\delta_{BC}E^{a_1\cdots a_l}(-)^{pl}\theta^{a_{p+1}}\cdots\theta^{a_{l}}
\frac{d}{d\theta^{a_p}}\cdots\frac{d}{d\theta^{a_1}}\psi^A(\theta)
\eea
Finally we reorder the indices of $E$ to match the ordering of the
$\theta$ and $d/d\theta$ factors
\bea
E^{a_1\cdots a_l}=(-)^{p(l-p)+p(p-1)/2}E^{a_{p+1}\cdots a_la_p\cdots a_1}
\eea
and then rename them to get
\bea
{}[\Omega^{AB,p}_{E_l},{\bar\psi}^C(\theta)]
&=&\delta_{BC}E^{a_1\cdots a_l}(-)^{p(p+1)/2}\theta^{a_{1}}\cdots
\theta^{a_{l-p}}
\frac{d}{d\theta^{a_{l-p+1}}}\cdots\frac{d}{d\theta^{a_l}}{\bar\psi}^A(\theta)
\eea
A word about suppressed color indices in this equation. The color indices
can all be left uncontracted. In that case there is a suppressed Kronecker
delta $\delta_\alpha^\delta\delta_\gamma^\beta$ on the right side
which requires the color indices of $({\bar\psi}^C)_\alpha^\beta$
to match the color indices of the factor $(\phi^B)_\gamma^\delta$ in $\Omega^{AB}$.
Then the color indices
of the factor ${\bar\phi}^A$ are identical to those of the ${\bar\psi}^A$
on the right side. With this in mind one can choose
to do any number of contractions on the color indices without
spoiling the validity of the equation.

Now one finds, for the 
supersymmetric combination
\bea
{}[\Omega^{AB}_{E_l},{\bar\psi}^C(\theta)]
&=&\gamma^l\delta_{BC}E^{a_1\cdots a_l}\sum_{p=0}^l\left(
\frac{\gamma^*}{\gamma}\right)^p{l\choose p}\theta^{a_{1}}\cdots
\theta^{a_{l-p}}
\frac{d}{d\theta^{a_{l-p+1}}}\cdots\frac{d}{d\theta^{a_l}}{\bar\psi}^A(\theta)
\nonumber\\
&=&\gamma^l\delta_{BC}E^{a_1\cdots a_l}\left(\theta^{a_1}+\frac{\gamma^*}{\gamma}
\frac{d}{d\theta^{a_1}}\right)\cdots\left(\theta^{a_l}+\frac{\gamma^*}{\gamma}
\frac{d}{d\theta^{a_l}}\right){\bar\psi}^A(\theta)\nonumber\\
&=&\delta_{BC}E^{a_1\cdots a_l}\left(\gamma\theta^{a_1}+{\gamma^*}
\frac{d}{d\theta^{a_1}}\right)\cdots\left(\gamma\theta^{a_l}+{\gamma^*}
\frac{d}{d\theta^{a_l}}\right){\bar\psi}^A(\theta)
\label{omegaonpsi}
\eea
A quick check on this final answer is to note that the action of
$Q^a$ on ${\bar\psi}^C$ is given by
\bea
{}[Q^a,{\bar\psi}^C]_\pm&=&\left(\gamma\theta^{a}-{\gamma^*}
\frac{d}{d\theta^{a}}\right){\bar\psi}^C
\eea
and the differential operator on the right anticommutes with each of the
factor operators in the action of $\Omega^{AB}_{E_l}$:
\bea
\left\{\gamma\theta^{a}-{\gamma^*}
\frac{d}{d\theta^{a}},\gamma\theta^{b}+{\gamma^*}
\frac{d}{d\theta^{b}} \right\}&=&0,
\eea
which is necessarily true if $\Omega^{AB}_{E_l}$ is indeed a supersymmetric
operator.

We have shown that $\Omega^{AB}_{E_l}$ can produce the insertion operators
required for the IIB superstring. It remains to use them
to construct color singlet terms to be added to the Hamiltonian such that
they don't contribute at leading order as $N\to\infty$,
but they do contribute at order $1/N$ in such a way that
the necessary operators accompany each closed string fission or fusion
transition. The way to do this is to construct color singlet quartic operators
with the schematic  color structure
\bea
\frac{1}{N}\Tr :{\bar\phi}\phi{\bar\phi}\phi:
=\frac{1}{N}:({\bar\phi})_\alpha^\beta(\phi)_\beta^\gamma
({\bar\phi})_\gamma^\delta(\phi)_\delta^\alpha :
\eea
where the color indices are contracted as in matrix multiplication.
The normal ordering removes the $O(1)$ contraction, which would arise
if the normal ordering were not specified. The terms dropped by
normal ordering would be supersymmetric bilinears which could be
included independently of the quartic terms, if desired.

Acting on multi-trace states this operator doesn't contribute at leading
order because contraction of the two annihilation operators will
not produce a factor of $N$ to compensate the $1/N$ out front. Furthermore
any such contraction will change the trace structure of the state,
splitting a trace into two traces, or joining
two traces into one.

To build the desired  terms we first contract two of the color indices of
$\Omega^{AB}_{E_l}$ to form a matrix  
${\overrightarrow \Omega}^{AB}_{E_l}$. Then we multiply this operator
with ${\overrightarrow\rho}^{CD}$ and take the trace to form the color singlet
  \bea
  \frac{1}{N}  \Tr :{\overrightarrow \Omega}^{AB}_{E_l}
  {\overrightarrow\rho}^{CD}:
  \eea
  By construction this is a supersymmetric color singlet operator
  with the appropriate color contraction scheme to first contribute
  at order $1/N$ in the large $N$ expansion. We confirm this by applying
  it to a multi-trace state. On a single trace state we have,
  consulting (\ref{omegaonpsi})
  \bea
  &&\frac{1}{N}  \Tr :{\overrightarrow \Omega}^{AB}_{E_l}
  {\overrightarrow\rho}^{CD}:T(A_1\theta_1;\cdots;A_M\theta_M)\ket{0}
  =\frac{1}{N}\sum_{k,m} \left[\delta_{DA_k}\delta_{BA_m};
    T(C\theta_k;\cdots;A_{m-1}\theta_{m-1})\right.\nonumber\\
 &&\hskip2in \left. E^{a_1\cdots a_l}\Lambda^{a_1}_m\cdots\Lambda^{a_l}_m
    T(A\theta_m;A_{m+1}\cdots
    ;A_{k-1}\theta_{k-1})\right]\ket{0}
  \eea
  where
  \bea
  \Lambda_m^a=\gamma\theta_m^a+\gamma^*\frac{d}{d\theta_m^a}
  \eea
  We see that the action has split the trace into two traces and the
  $1/N$ factor is uncanceled. For $l=0,2,4,6,8$ the action of this operator
  produces the five insertions
  listed in (\ref{insertions}) respectively.

  The fusion of two traces into one is also produced by this operator
  acting on a state containing two or more traces. This happens when
  the two annihilation operators contract against distinct traces.
  \bea
  \frac{1}{N}\Tr :{\overrightarrow \Omega}^{AB}_{E_l}
  {\overrightarrow\rho}^{CD}:T({A_1\theta_1\cdots A_K\theta_K})
T({B_1\eta_{1}\cdots B_L\eta_L})\ket{0}_{\rm Fusion}&=&\label{fuse2}
\\
&&\hskip-4.5in +\frac{1}{N}\sum_{k=1}^K
\sum_{m=1}^L\delta_{DA_k}\delta_{BB_m} 
E^{a_1\cdots a_l}\Lambda^{a_1}_{\eta_m}\cdots\Lambda^{a_l}_{\eta_m}T({C\theta_{k}
\cdots A_{k-1}\theta_{k-1}}A\eta_m\cdots B_{m-1}\eta_{m-1})\ket{0}
\nonumber\\
&&\hskip-4.5 in +\frac{1}{N}\sum_{k=1}^K\sum_{l=1}^L\delta_{DB_l}\delta_{BA_k}
E^{a_1\cdots a_l}\Lambda^{a_1}_{\theta_k}\cdots\Lambda^{a_l}_{\theta_k}
T(A\theta_k\cdots A_{k-1}\theta_{k-1}C{\eta_{m}
\cdots B_{m-1}\eta_{m-1}})\ket{0}.\nonumber
\eea
If supersymmetry were the only requirement for modifying the Hamiltonian,
We could form $J_{ABCD}\Tr :{\overrightarrow \Omega}^{AB}_{E_l}
{\overrightarrow\rho}^{CD}:$ with $J_{ABCD}$ arbitrary
complex numbers  and add it plus its hermitian conjugate to the
Hamiltonian. A simple choice for $J$ which works for the IIB
superstring is $J_{ABCD}=\delta_{AB}\delta_{CD}$.

For the IIB superstring, the antisymmetric tensors $E_l$ are built from
$SO(8)$ gamma matrices \cite{greenschwarzbrink}. For even $l$, as required
in (\ref{insertions}),
they are second rank tensors in $SO(8)$ vector indices, $E^{kl}_{2l}$, which
can read off from (\ref{insertions}). The relative coefficients in the
complete insertion operator cannot
be taken directly from \cite{greenschwarzbrink},
because in the latter reference the continuum limit has already
been partially
taken, holding the insertion a fixed distance from the interaction point.
The string bit approach is discrete at a fundamental level, and the continuum
limit is properly regarded
as a low excitation energy approximation on chains with large numbers
of bits, which must be carefully analyzed to make a detailed comparison.
The terms we have constructed will prescribe that the insertion is
within one or two discrete units away from the location of the separation or
joining point, whereas the approach of \cite{greenschwarzbrink} would
correspond to placing the insertion
$\epsilon M$, with $\epsilon$ small but fixed,
units away. Such a prescription would be
far from natural in the string bit approach. The continuum
limits in these two methods are not taken in the same way. Although
it seems plausible
that term by term the two methods will give results that are proportional,
it would be premature
to expect that the proportionality constants be exactly one. 
\section{Conclusion}
We have found many ways to modify the string bit
Hamiltonian of \cite{thornspace} while maintaining
supersymmetry. In particular, we have constructed
explicit supersymmetric operators which, added to the
Hamiltonian, can provide the insertion prefactors
necessary for the overlap prescription for lightcone superstring
vertices to give correct scattering amplitudes. This ties up one of
the loose ends in \cite{thornspace}. In this discussion we have
treated the coordinate world sheet fields as fundamental rather
than as effective fields simulating low energy excitations of a
discrete flavor variable. In \cite{thornspace} we showed that a system of
Heisenberg chains described by such an effective field can be obtained in the
$N\to\infty$ limit from the dynamics of $2^d$ valued ``flavor'' indices.
It will be interesting to pursue this idea further by extending it
to interactions. Specifically, understanding how the ${\tilde X}^k X^l$ factor in the
prefactor (\ref{fullinsertion}) can be obtained from such a discrete
flavor dynamics is a natural next step.
\vskip14pt
\noindent\underline{Acknowledgments}: I would like to thank Ethan Torres
for his early contributions to this project and for valuable discussions.
This research was supported in part by the Department
of Energy under Grant No. DE-SC0010296. 

\appendix
\section{Proof of Eq~(\ref{Qanticom})}
\bea
{}[Q^c,\Omega^p_{a_1\cdots a_l}]&=&\sum_k\frac{(-)^{lk}}{k!}\bigg[
(-)^{k+p}\gamma^*{\bar\phi}_{ca_1\cdots a_pb_1\cdots b_k}
\phi_{a_{p+1}\cdots a_lb_1\cdots b_k}\nonumber\\
&&+\gamma\sum_{m=1}^p(-)^{m+k+p}\delta_{ca_m}{\bar\phi}_{a_1\cdots{\hat a}_m
  \cdots a_pb_1\cdots b_k}
\phi_{a_{p+1}\cdots a_lb_1\cdots b_k}\nonumber\\
&&+\gamma\sum_{m=1}^k(-)^{m+k+1}\delta_{cb_m}{\bar\phi}_{a_1\cdots a_pb_1
  \cdots{\hat b}_m\cdots b_k}
\phi_{a_{p+1}\cdots a_lb_1\cdots  b_k}\bigg]\nonumber\\
&&+\sum_k\frac{(-)^{lk}}{k!}\bigg[
(-)^{k+p+1}\gamma {\bar\phi}_{a_1\cdots a_pb_1\cdots b_k}
\phi_{ca_{p+1}\cdots a_lb_1\cdots b_k}\nonumber\\
&&+\gamma^*\sum_{m=1}^{l-p}(-)^{m+k+p+1}\delta_{ca_{m+p}}
{\bar\phi}_{a_1
  \cdots a_pb_1\cdots b_k}
\phi_{a_{p+1}\cdots  {\hat a}_{m+p}\cdots  a_lb_1\cdots b_k}\nonumber\\
&&+\gamma^*\sum_{m=1}^k(-)^{m+k+l+1}\delta_{cb_m}{\bar\phi}_{a_1\cdots a_pb_1
  \cdots b_k}
\phi_{a_{p+1}\cdots a_lb_1\cdots{\hat b}_m\cdots b_k}\bigg]
\eea
The first three lines of this equation arise from the bracket of
$Q$ with ${\bar\phi}$ and the last three lines from that of
$Q$ with $\phi$.

The first line is cancelled by the sixth line. To see this
use the Kronecker delta of the latter to set $b_m= c$,
and then use the antisymmetry in indices.
Then the summand is independent of $m$, so the sum over $m$
just gives a factor of $k$, which changes $1/k!\to 1/(k-1)!$.
Then shifting $k\to k+1$ gives
$-$ line 1. Similarly line 3 cancels line 4. Lines 2 and 5 are all that's
left: 
\bea
{}[Q^c,\Omega^p_{a_1\cdots a_l}]&=&\sum_k\frac{(-)^{lk}}{k!}\bigg[
+\gamma\sum_{m=1}^p(-)^{m+k+p}\delta_{ca_m}{\bar\phi}_{a_1\cdots{\hat a}_m
  \cdots a_pb_1\cdots b_k}
\phi_{a_{p+1}\cdots a_lb_1\cdots b_k}\nonumber\\
&&+\gamma^*\sum_{m=1}^{l-p}(-)^{m+k+p+1}\delta_{ca_{m+p}}
{\bar\phi}_{a_1
  \cdots a_pb_1\cdots b_k}
\phi_{a_{p+1}\cdots  {\hat a}_{m+p}\cdots  a_lb_1\cdots b_k}
\bigg]
\eea
The two sums over $k$ can now be recognized as two of the $\Omega$'s:
\bea
{}[Q^c,\Omega^p_{a_1\cdots a_l}]&=&
+\gamma\sum_{m=1}^p(-)^{m+p}\delta_{ca_m}\Omega^{p-1}_{a_1\cdots{\hat a}_m
  \cdots a_l}+\gamma^*\sum_{m=p+1}^{l}(-)^{m+1}
\delta_{ca_{m}}\Omega^{p}_{a_1\cdots{\hat a}_m
  \cdots a_l}
\eea
which is (\ref{Qanticom})

\end{document}